\documentclass[longauth]{aa} 

\usepackage{graphicx}
\usepackage{rotating}
\usepackage{longtable}
\usepackage[varg]{txfonts}
\usepackage[T1]{fontenc}
\usepackage[utf8]{inputenc}
\usepackage{orcidlink}

\begin{document} 

   \title{A long-lasting eruption heralds SN~2023ldh, a clone of SN~2009ip}

   \author{A.~Pastorello\inst{1}\fnmsep\thanks{andrea.pastorello@inaf.it}\orcidlink{0000-0002-7259-4624}
          \and
          A.~Reguitti\inst{2,1}\orcidlink{0000-0003-4254-2724}
          \and 
          L.~Tartaglia\inst{3}\orcidlink{0000-0003-3433-1492}
           \and
          G.~Valerin\inst{1}\orcidlink{0000-0002-3334-4585}
          \and 
          Y.-Z.~Cai\inst{4,5,6}\orcidlink{0000-0002-7714-493X}
          \and
          P.~Charalampopoulos\inst{7}\orcidlink{0000-0002-0326-6715}
          \and
          F.~De~Luise\inst{3}\orcidlink{0000-0002-6570-8208}          
          \and 
          Y.~Dong\inst{8,9}\orcidlink{0000-0002-7937-6371}
          \and
          N.~Elias-Rosa\inst{1,10}\orcidlink{0000-0002-1381-9125} 
          \and
          J.~Farah\inst{11,12}\orcidlink{0000-0003-4914-5625}
          \and
          A.~Farina\inst{13}\orcidlink{0009-0005-5257-8319}
          \and
          S.~Fiscale\inst{14}\orcidlink{0000-0001-8371-8525}
          \and
          M.~Fraser\inst{15}\orcidlink{0000-0003-2191-1674}
          \and
          L.~Galbany\inst{10,16}\orcidlink{0000-0002-1296-6887}
          \and
          S.~Gomez\inst{9}\orcidlink{0000-0001-6395-6702}
          \and
          M.~Gonz\'alez-Ba\~nuelos\inst{10,16}\orcidlink{0009-0006-6238-3598} 
          \and
          D.~Hiramatsu\inst{9,17}\orcidlink{0000-0002-1125-9187}
          \and
          D.~A.~Howell\inst{11,12}\orcidlink{0000-0003-4253-656X}
          \and
          T.~Kangas\inst{7,18}\orcidlink{0000-0002-5477-0217}
          \and
          T.~L.~Killestein\inst{7}\orcidlink{0000-0002-0440-9597}
          \and
          P.~Marziani\inst{1}\orcidlink{0000-0002-6058-4912}
          \and
          P.~A.~Mazzali\inst{19,20}\orcidlink{0000-0001-6876-8284}
          \and 
          E.~Mazzotta Epifani\inst{21}\orcidlink{0000-0003-1412-0946}
          \and
          C.~McCully\inst{11}\orcidlink{0000-0001-5807-7893}
          \and
          P.~Ochner\inst{13,1}\orcidlink{0000-0001-5578-8614}
          \and
          E.~Padilla Gonzalez\inst{11,12,22}\orcidlink{0000-0003-0209-9246}
          \and
          A.~P.~Ravi\inst{8,9}\orcidlink{0000-0002-7937-6371}
          \and
          I.~Salmaso\inst{1,23}\orcidlink{0000-0003-1450-0869}
          \and
          S. Schuldt\inst{24,25}\orcidlink{0000-0003-2497-6334}
          \and
          A.~G.~Schweinfurth\inst{20,26}\orcidlink{0000-0002-8274-7196}
          \and
          S.~J.~Smartt\inst{27,28}\orcidlink{0000-0002-8229-1731}
          \and
          K.~W.~Smith\inst{27,28}\orcidlink{0000-0001-9535-3199}
          \and
          S.~Srivastav\inst{27}\orcidlink{0000-0003-4524-6883}
          \and
          M.~D.~Stritzinger\inst{29}\orcidlink{0000-0002-5571-1833}
          \and
          S.~Taubenberger\inst{26,20}\orcidlink{0000-0002-4265-1958}
          \and
          G.~Terreran\inst{30,11,12}\orcidlink{0000-0003-0794-5982}
          \and
          S.~Valenti\inst{8}\orcidlink{0000-0001-8818-0795}
          \and
          Z.-Y.~Wang\inst{31,32}\orcidlink{0000-0002-0025-0179}
          \and
          F.~Guidolin\inst{2,33}\orcidlink{0009-0001-1215-9380}
          \and
          C.~P.~Guti\'errez\inst{10,16}\orcidlink{0000-0003-2375-2064}
          \and
          K.~Itagaki\inst{34}
          \and
          S.~Kiyota\inst{35}
          \and
          P. Lundqvist \inst{36}\orcidlink{0000-0002-3664-8082}
          \and
          K.~C. Chambers\inst{37}\orcidlink{0000-0001-6965-7789}
          \and
          T.~J.~L.~de~Boer\inst{37}\orcidlink{0000-0001-5486-2747}
          \and
          C.-C.~Lin\inst{37}\orcidlink{0000-0002-7272-5129}
          \and
          T.~B.~Lowe\inst{37}\orcidlink{0000-0002-9438-3617}
          \and
          E.A.~Magnier\inst{37}\orcidlink{0000-0002-7965-2815}
          \and
          R.~J.~Wainscoat\inst{37}\orcidlink{0000-0002-1341-0952}
          }

   \institute{INAF - Osservatorio Astronomico di Padova, Vicolo dell'Osservatorio 5, I-35122 Padova, Italy \\   \email{andrea.pastorello@inaf.it} 
         \and
          INAF – Osservatorio Astronomico di Brera, via E. Bianchi 46 I-23807 Merate, Italy 
         \and
          INAF – Osservatorio Astronomico d'Abruzzo, via Mentore Maggini, I-64100 Teramo, Italy 
           \and
           Yunnan Observatories, Chinese Academy of Sciences, Kunming 650216, P.R. China 
           \and
           Key Laboratory for the Structure and Evolution of Celestial Objects, Chinese Academy of Sciences, Kunming 650216, P.R. China 
           \and
        International Centre of Supernovae, Yunnan Key Laboratory, Kunming 650216, P.R. China 
           \and
         Department of Physics and Astronomy, University of Turku, Vesilinnantie 5, FI-20014 Turku, Finland 
         \and
          Department of Physics and Astronomy, University of California, 1 Shields Avenue, Davis, CA 95616-5270, USA  
         \and
         Center for Astrophysics \textbar{} Harvard $\&$ Smithsonian, 60 Garden Street, Cambridge, MA 02138-1516, USA 
         \and
          Institute of Space Sciences (ICE, CSIC), Campus UAB, Carrer de Can Magrans s/n, E-08193, Barcelona, Spain 
         \and
          Las Cumbres Observatory, 6740 Cortona Dr \#102, Goleta, CA 93117, USA  
         \and
        Department of Physics, University of California, Santa Barbara, Santa Barbara, CA 93106, USA  
          \and
          Universit\'a degli Studi di Padova, Dipartimento di Fisica e Astronomia, Vicolo dell’Osservatorio 2, 35122 Padova, Italy 
         \and
          Science and Technology Department, Parthenope University of Naples, Centro Direzionale, Isola C4, I-80143 Naples, Italy           
         \and
         School of Physics, O’Brien Centre for Science North, University College Dublin, Belfield, Dublin 4, Ireland  
         \and
          Istitut d'Estudis Espacials de Catalunya (IEEC), E-08860 Castelldefels (Barcelona), Spain 
          \and  
          The NSF AI Institute for Artificial Intelligence and Fundamental Interactions, USA 
         \and
         Finnish Centre for Astronomy with ESO (FINCA), University of Turku, 20014 Turku, Finland  
         \and
         Astrophysics Research Institute, Liverpool John Moores University, IC2, 146 Brownlow Hill, Liverpool L3 5RF, UK         
         \and
          Max Planck Institute for Astrophysics, Karl-Schwarzschild-Str. 1, 85741 Garching, Germany 
         \and 
          INAF – Osservatorio Astronomico di Roma, Via Frascati 33, I-00078, Monte Porzio Catone, Italy  
         \and
          Johns Hopkins University, 3400 N. Charles Street, Baltimore, MD 21218, USA 
         \and
         INAF-Osservatorio Astronomico di Capodimonte, Via Moiariello 16, 80131 Napoli, Italy   
         \and
         Dipartimento di Fisica, Universit\`a  degli Studi di Milano, via Celoria 16, I-20133 Milano, Italy
         \and
         INAF - IASF Milano, via A. Corti 12, I-20133 Milano, Italy
         \and
          Technical University of Munich, TUM School of Natural Sciences, Physics Department, James-Franck-Str. 1, 85741 Garching, Germany 
          \and
          Department of Physics, University of Oxford, Denys Wikinson Building, Kable Road, Oxford OX1 3RH, UK 
         \and
          Astrophysics Research Centre, School of Mathematics and Physics, Queen’s University Belfast, BT7 1NN, UK 
         \and
          Department of Physics and Astronomy, Aarhus University, Ny Munkegade 120, 8000 Aarhus C, Denmark 
        \and
         Adler Planetarium, 1300 S. DuSable Lake Shore Drive, Chicago, IL 60605, USA
         \and
         School of Physics and Astronomy, Beijing Normal University, Beijing 100875, China
         \and
         Department of Physics, Faculty of Arts and Sciences, Beijing Normal University, Zhuhai 519087, China         
         \and
          Universit\'a degli Studi dell'Insubria, Dipartimento di Scienza e Alta Tecnologia, Via Valeggio 11, I-22100, Como, Italy
         \and
          Itagaki Astronomical Observatory, Yamagata 990-2492, Japan      
         \and
         Variable Star Observers League in Japan, 7-1 Kitahatsutomi, Kamagaya, Chiba 273-0126, Japan 
         \and
         The Oskar Klein Centre, Department of Astronomy, Stockholm University, AlbaNova SE-10691, Stockholm, Sweden
         \and
         Institute for Astronomy, University of Hawaii, 2680 Woodlawn Drive, Honolulu HI 96822, USA
}

   \date{Received 29 March 2025 / Accepted 4 July 2025}

\abstract {We discuss the results of the spectroscopic and photometric monitoring of the type IIn supernova (SN)~2023ldh. Survey archive data show that the SN progenitor experienced erratic variability in the years before exploding.
From May 2023, the source shows a general slow luminosity rise lasting over four months with some superposed luminosity fluctuations. In analogy to \object{SN~2009ip}, we label this brightening as Event~A. 
During Event~A, \object{SN~2023ldh} reaches a maximum absolute magnitude of $M_r = -15.52 \pm 0.24$ mag. Then the light curves show a luminosity decline of about 1 mag in all filters lasting about two weeks, followed by a steep brightening (Event~B) to an absolute peak magnitude of $M_r = -18.53 \pm 0.23$ mag, replicating the evolution of \object{SN~2009ip} and similar SNe~IIn.
Three spectra of \object{SN~2023ldh} are obtained during Event~A, showing multi-component P~Cygni profiles of H~I and Fe~II lines. 
During the rise to the Event~B peak, the spectrum shows a blue continuum dominated by Balmer lines in emission with Lorentzian profiles, with a full width at half-maximum (FWHM) velocity of about 650 km s$^{-1}$.
Later, in the post-peak phase, the spectrum reddens, and broader wings appear in the 
 H$\alpha$ line profile. Metal lines are well visible with P~Cygni profiles and velocities of about 2000 km s$^{-1}$. Beginning around three months past maximum and until very late phases, the Ca II lines become among the most prominent features, while H$\alpha$ is dominated by an intermediate-width component with a boxy profile. Although \object{SN~2023ldh} mimics the evolution of other \object{SN~2009ip}-like transients, it is slightly more luminous and has a slower photometric evolution. The surprisingly homogeneous observational properties of \object{SN~2009ip}-like events may indicate similar explosion scenarios and similar progenitor parameters.}

   \keywords{supernovae: general -- supernovae: individual: SN~2023ldh -- supernovae: individual: SN~2009ip -- Stars: winds, outflows}
   \maketitle

\section{Introduction} \label{sect:intro}

The latest stages of life of massive stars are poorly known and, in some cases, the observations seem to contradict the theoretical predictions. Unequivocal observational evidence indicates that a massive star may experience recurrent
short-duration outbursts, or even longer-lasting eruptive phases a short time before finally exploding as a supernova \citep[SN; for a review, see][]{smi14}. 
In particular, clear signatures of pre-SN variability are frequently observed in the progenitors of supernovae
(SNe) showing interaction with circumstellar material (CSM). Although major outbursts are frequently observed in association with H-rich progenitors of type IIn SNe \cite[e.g.,][and references therein]{ofe14,str21,reg24}, pre-SN variability
in H-poor, He-rich progenitors of SNe Ibn is only occasionally observed \citep[e.g.,][]{pasto07,bre24}.

Spectroscopic monitoring of the pre-SN eruptive phase is available in a handful of SNe IIn:  \object{SN~2009ip} \citep{smi10,fol11,pasto13,mau13},
\object{SN~2015bh} \citep{nancy16,tho17}, \object{SN~2016jbu} \citep{kil17,bre22a}. More recently, pre-SN spectra were also obtained for the type Ibn \object{SN~2023fyq}  \citep{bre24,dong24}.
The combination of photometric monitoring and spectroscopic observations obtained before the stellar core collapse is fundamental for the characterization of the progenitor star.
For this reason, it is crucial to obtain spectral sequences for all species of transients sitting in the luminosity gap that separates
classical novae and core-collapse supernovae \citep{kas12,pasto19,cai22}, as a fraction of them have been seen to explode as a SN a short time later. In this paper, we discuss the case of an object initially identified as a long-lasting gap transient that later experienced a major brightening compatible with a type IIn SN explosion: \object{SN~ 2023ldh}. A combination of long-lasting photometric monitoring
and spectra from the pre-SN eruptive phase to the late nebular phase makes \object{SN~2023ldh} one of the interacting SNe with the most comprehensive datasets.

The optical transient \object{SN~2023ldh}\footnote{The object is also known with the following survey designations: ZTF23aamanim, ATLAS23qiw and PS23hmg.} was discovered by the Zwicky Transient Facility \citep[ZTF,][]{bel19,gra19} on 2023 May 28.32 UT, at a $g$-band magnitude of 20.74 \citep{de23}.
The coordinates are: RA = 15:09:09.597 and Dec = +52:31:59.80 (J2000), $32''.2$ west and $17''.6$ north of the centre of the spiral galaxy (SAb-type) \object{NGC~5875}.
The location of \object{SN~2023ldh} is nearly coincident with an extended source, \object{SDSS J150909.51+523159.0}, and is also very close (about 0.2 arcmin) to the explosion site of the Ca-rich transient \object{SN~2022oqm} \citep{sol22,zim22,ful22,ira24,yad24}.
The closest source within \object{NGC~5875} with known redshift (z $= 0.011321 \pm 0.000016$) is \object{SDSS J150912.08+523154.3}, only $\sim$0.39 arcmin south-east of \object{SN~2023ldh}.

Later, on 2023 October 13.46 UT, K. Itagaki announced a brightening of the source to an unfiltered magnitude of 16, and a low signal-to-noise spectrum was obtained with the 1.82-m Copernico Telescope of the Asiago Observatory (Italy), allowing us to classify the object as an SN IIn \citep{pasto23}.

\begin{table*}
\caption{\label{photpeak} Timing (MJDs), apparent  and absolute magnitudes at maximum in the different bands, as obtained from fitting
the light curve with a low-order polynomial. 
}
\centering
\begin{tabular}{ccccccc}
\hline\hline
Band & MJD$_{A}^{peak}$ & $m_{A}^{peak}$ & $M_{A}^{peak}$ & MJD$_{B}^{peak}$ & $m_{B}^{peak}$ & $M_{B}^{peak}$ \\
\hline
B (Vega) &          --     &       --       &       --         & 60235.8 $\pm$ 0.5 & 16.00 $\pm$ 0.03 & -18.59 $\pm$ 0.27 \\
V (Vega) & 60201.5 $\pm$ 1.0 & 18.77 $\pm$ 0.04 &  -15.59 $\pm$ 0.25 & 60237.0 $\pm$ 0.3 & 15.79 $\pm$ 0.03 & -18.57 $\pm$ 0.25 \\
u (ab)   &          --     &       --       &       --         & 60234.6 $\pm$ 2.1 & 16.19 $\pm$ 0.04 & -18.55 $\pm$ 0.30 \\
g (ab)   & 60200.3 $\pm$ 3.4 & 19.09 $\pm$ 0.14 &  -15.42 $\pm$ 0.30 & 60236.3 $\pm$ 1.5 & 15.76 $\pm$ 0.02 & -18.75 $\pm$ 0.26 \\
r (ab)   & 60200.6 $\pm$ 1.7 & 18.75 $\pm$ 0.06 &  -15.52 $\pm$ 0.24 & 60237.0 $\pm$ 0.4 & 15.74 $\pm$ 0.02 & -18.53 $\pm$ 0.23 \\
i (ab)   & 60202.6 $\pm$ 2.7 & 18.89 $\pm$ 0.09 &  -15.25 $\pm$ 0.24 & 60237.8 $\pm$ 0.4 & 15.92 $\pm$ 0.01 & -18.22 $\pm$ 0.22 \\
z (ab)   &          --     &       --       &       --         & 60238.0 $\pm$ 1.0 & 16.01 $\pm$ 0.04 & -18.00 $\pm$ 0.21 \\
c (ab)   & 60200.5 $\pm$ 3.4 & 19.10 $\pm$ 0.10 &  -15.29 $\pm$ 0.27 & 60237.2 $\pm$ 0.9 & 15.87 $\pm$ 0.07 & -18.52 $\pm$ 0.26 \\ 
o (ab)   & 60201.9 $\pm$ 2.1 & 18.84 $\pm$ 0.11 &  -15.37 $\pm$ 0.25 & 60237.8 $\pm$ 1.2 & 15.84 $\pm$ 0.09 & -18.37 $\pm$ 0.24 \\ \hline
\end{tabular}
\end{table*}

The distance of the host galaxy was estimated using the Tully-Fisher method. We adopt the most recent estimate of \citet{tul13},
reported to the following cosmological parameters: the Hubble constant H$_0$ = 73~km~s$^{-1}$~Mpc$^{-1}$, $\Omega_{\rm matter} = 0.27$
and $\Omega_{\rm vacuum} = 0.73$. With a host galaxy distance of $54.5 \pm 5.0$ Mpc, we infer a distance modulus of $\mu = 33.68 \pm 0.20$ mag.\footnote{This Tully-Fisher distance to NGC~5875 is consistent with the kinematic distance (after corrections for the influence of the Virgo Cluster, the Great Attractor, and the Shapley Supercluster); adopting the same cosmologic parameters as above, we obtain $\mu = 33.70 \pm 0.15$ mag \citep{mou00}.}

The interstellar dust reddening towards \object{SN~2023ldh} is composed by a modest Milky Way contribution plus a significant component within the 
host galaxy. While we can infer the former component from \citet{sch11}, $E(B-V)_{MW}= 0.016$ mag, the latter can be obtained through the analysis of
our highest resolution spectrum of the SN (see Sect. \ref{sect:reddening} for a discussion on the method).  
Including the host galaxy contribution, we infer a total colour excess of $E(B-V)_{tot}= 0.216 \pm 0.044$ mag in the direction of \object{SN~2023ldh}. This value will be used throughout this paper.

   \begin{figure*}
   \centering
   \includegraphics[width=17.3cm]{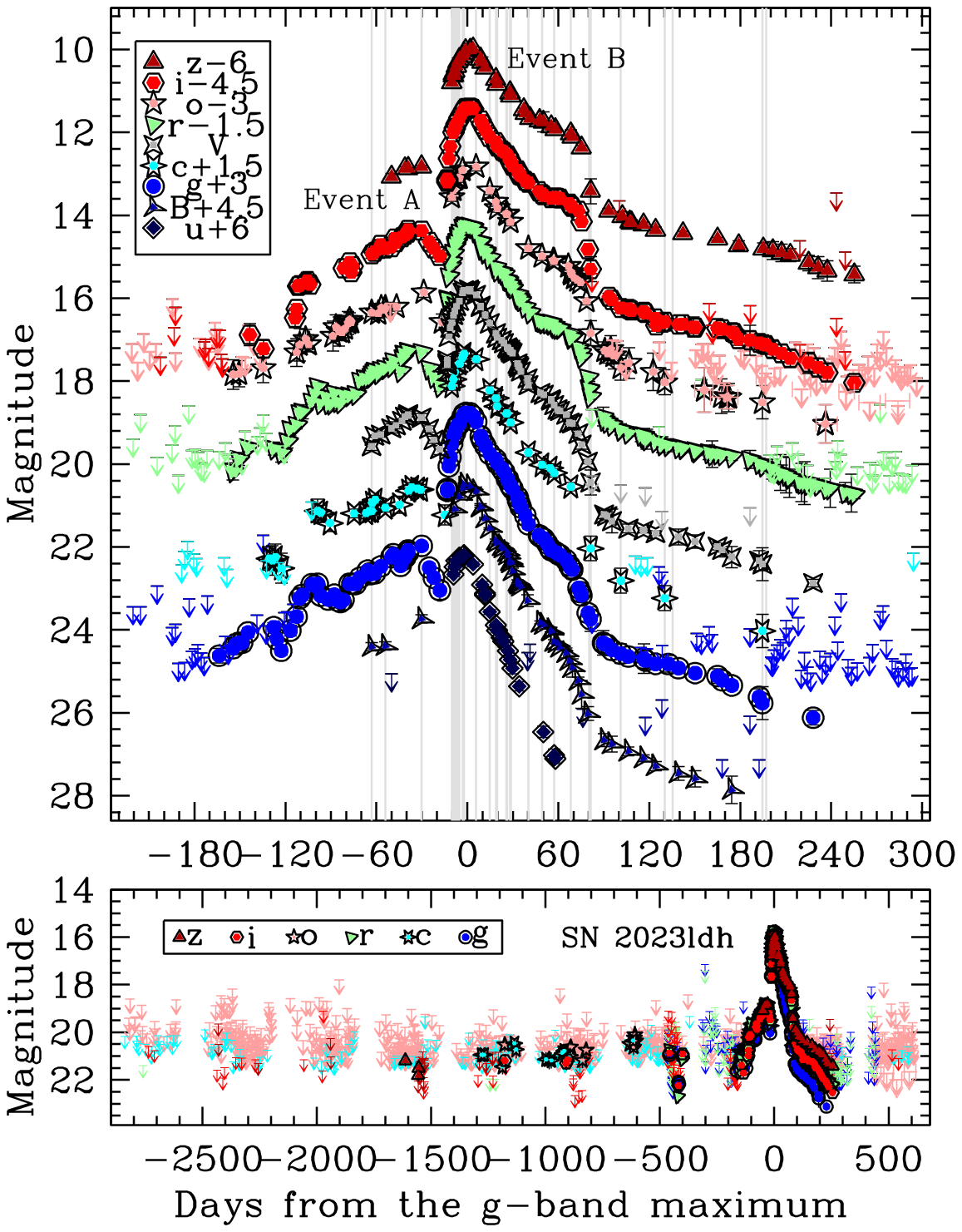}
      \caption{Observed multi-band light curves of \object{SN~2023ldh}. Top panel: blow-up of the light curve containing the two main brightening episodes:
Event A and Event B. Vertical thin grey lines mark the epochs of the spectroscopic observations. Bottom panel: Light curves of \object{SN~2023ldh} in a limited number of filters, spanning a period of about nine years. 
Johnson-Bessell data are in Vega magnitude scale, while Sloan and ATLAS data are in AB magnitude scale. The phases are computed from the time of
the $r$-band Event B maximum (MJD = 60237.0 $\pm$ 0.4).}
         \label{lightcurve}
   \end{figure*}

\section{Observations} \label{sect:obs}

A short time after the discovery announcement of \object{SN~2023ldh}, we initiated a monitoring campaign aimed to study the evolution  of the transient in the context of the Nordic Optical Telescope (NOT) Unbiased Transient Survey-2 (NUTS-2) program.\footnote{\url{https://nuts.sn.ie/}} 
For the photometric campaign, we made use of the following facilities: the 2m Liverpool Telescope (LT) equipped with IO:O, the 2.56m Nordic Optical Telescope 
(NOT) with the Alhambra Faint Object Spectrograph and Camera (ALFOSC), the 10.4m Gran Telescopio Canarias (GTC) with the Optical System for Imaging and low-Intermediate-Resolution Integrated Spectroscopy (OSIRIS+), hosted in La Palma (Spain); the 1.82m Copernico Telescope with the Asiago Faint Object Spectrograph and Camera (AFOSC) and the 67/91cm Schmidt Telescope with a Moravian detector, hosted at the Asiago Observatory, Mt. Ekar (Italy); the Wide-field Optical Telescope (WOT), a 68/91cm Schmidt Telescope equipped with an Apogee Aspen CCD, at the Campo Imperatore Observatory (Italy); the Joan Or\'o Telescope (TJO) with MEIA2 at the Montsec Astronomical Observatory (Spain); the 1m-class telescopes of Las Cumbres Observatory (LCO) telescope network\footnote{\url{https://lco.global/}} equipped with Sinistro cameras in the framework of the Global Supernova Project. Additional data were collected through 0.5m class amateur facilities of the iTelescope network\footnote {\url{https://support.itelescope.net/support/home}} hosted in Utah and California (USA), and others were obtained at Itagaki Observatory of Yamagata (Japan). Crucial pre-SN data were finally collected from the public surveys, including ZTF
\footnote{Images retrieved through the NASA/IPAC Infrared Science Archive: \url{https://irsa.ipac.caltech.edu/Missions/ztf.html}.} (in the $g$ and $r$ bands), ATLAS\footnote{The template-subtracted forced photometry is released through the Asteroid Terrestrial-impact Last Alert System (ATLAS) data-release interface: \url{https://fallingstar-data.com/forcedphot/queue/}.} \citep[with the cyan and orange filters -- hereafter $c$ and $o$;][]{ton18,smi20}, and Pan-STARRS \citep[in the $r$, $i$, and $z$ bands;][]{cha19,fle20,mag20}. Including the survey data, the photometric data analysed in this paper cover a period of about nine years, from 2016 to early 2025.

We also collected about thirty optical spectra (see Sect. \ref{sect:spec}), spanning a period of about 260 days. The spectra were obtained using GTC with OSIRIS+, the NOT with ALFOSC, the Copernico Telescope with AFOSC, the 3.58~m Telescopio Nazionale Galileo at La Palma (Spain) equipped with Dolores (LRS), the 6.5~m MMT telescope (Mount Hopkins, Arizona, USA) equipped with the Binospec spectrograph, and the 10~m Keck I telescope  with the Low Resolution Imaging Spectrometer (LRIS).

\section{Photometric data}\label{sect:photometricdata}

Photometric observations of \object{SN~2023ldh} were obtained with the following filters: Johnson-Bessell $B$ and $V$; Sloan $u$, $g$, $r$, $i$, and $z$; ATLAS  $o$ and $c$. While the ATLAS and Pan-STARRS magnitudes were directly collected through the survey data release interfaces (see Sect. \ref{sect:obs}), the reduction of ZTF data and the images obtained with our facilities was carried out using the {\sc SNOoPY} pipeline.\footnote{{\sc SNOoPY} is a package for supernova photometry using point-spread function (PSF) fitting and/or template subtraction developed by E. Cappellaro. A package description can be found at the website: \url{http://sngroup.oapd.inaf.it/ecsnoopy.html}.} After correcting the science frames for bias and flat-field images, we fine-tuned the astrometric calibration of the images and measured the instrumental PSF-fitting photometry of the target after the subtraction of the host galaxy template. \footnote{We used stacked Pan-STARRS images obtained before 2015 as templates in the Sloan $g$, $r$, $i$ and $z$ bands. For Johnson-Bessell $B$, $V$ and Sloan $u$ images, we used templates collected through the LCO archive, and obtained during the observational campaign of SN~2022oqn, when the magnitude of the precursor of SN~2023ldh was $g > 22$ mag.} The final photometric calibration was performed accounting for the zero point and colour-term corrections for each instrumental configuration, and making use of secondary standards from the Sloan Digital Sky Survey \citep[SDSS;][]{alm23} catalogue. To calibrate the photometric data in the Johnson-Cousins filters, a catalogue of comparison stars was obtained by converting Sloan magnitudes to Johnson-Cousins magnitudes using the transformation relations of \citet{cho08}. 

\subsection{Light curve evolution}\label{sect:lc}

The multiband optical light curves, from $-$230~d to about +300~d with respect to the maximum epoch of Event B , are shown in Fig. \ref{lightcurve} (top panel). All photometric data are reported in Table A1.\footnote{Table A1 is available in electronic form at the CDS, and contains the following information: the epoch and the MJD of the observation (Columns 1 and 2, respectively); the filter (Column 3); the magnitude and the error (Columns 4 and 5, respectively); the instrumental configuration (Column 6), additional information (Column 7).} Observations obtained by the surveys mentioned above since 2016 are also available, and are shown in the bottom panel of Fig. \ref{lightcurve} for a limited number of filters.

Earlier observations, obtained from mid-2016 by the main surveys or through the inspection of the public archives, show no sources to a limiting magnitude of about 22 in the $r$ band (see Table A.1).
Subsequent observations return in most cases detection limits, although a faint source is detected ($\gtrsim 3\sigma$) in the SN location at several epochs. In particular, the source is visible in $z$-band Pan-STARRS images obtained in mid-2019 (from May 20 to July 18), at magnitudes ranging from $z \sim 21.2$ to $21.8$ mag. 
In $i$-band Pan-STARRS and $c,o$ ATLAS images taken from April to September 2020 and from January to July 2021, the object is seen always at magnitudes fainter than $20.5$ mag.
In January and February 2022 the source is still detected in ATLAS images at slightly brighter magnitudes ($c \sim o \approx 20.4 \pm 0.2$ mag).
The source is also found in archival LCOGT $g,r,i$ images obtained in mid-2022 (on July 18, August 27, and September 13) with magnitudes varying erratically in the three filters (from $20.7$ to $22.7$ mag in the $r$ band). These luminosity fluctuations are produced by the same stellar system that will later generate SN~2023ldh, and are similar to the extreme variability expected in massive binaries containing luminous blue variables \citep[LBVs; e.g.,][]{wag04,pasto10,mue23,agh23a,agh23b,pus25}.

The transient is observed again from 2023 May 9, about 20 days before the official discovery date \citep{de23}. In this phase, the light curve of the transient shows a slow increase in luminosity (by about 2.5-3 magnitudes) in all optical filters lasting about four months. Significant luminosity fluctuations are observed superimposed on the global luminosity rise, in particular at about 1.5 months after the discovery. The average rise rates are $-1.9 \pm 0.1$ mag~(100~d)$^{-1}$ in the $g$ band, $-2.3 \pm 0.1$ mag~(100~d)$^{-1}$ in the $r$ band, and $-2.2 \pm 0.1$ mag~(100~d)$^{-1}$ in the $i$ band.
The epoch and the magnitude of the $r$-band light-curve maximum (MJD~$= 60200.6 \pm 1.7$ and $r = 18.75 \pm 0.06$ mag, respectively) are determined through a low-order polynomial fit. This first peak is reached about four months after the discovery, at a similar time for all filters (see Table \ref{photpeak}). Accounting for the line-of-sight extinction and the adopted host galaxy distance, we obtain the following absolute magnitude at the time of this first light-curve peak: $M_r = -15.52 \pm 0.24$ mag.\footnote{The error on the absolute magnitude is largely dominated by the uncertainty on the host galaxy distance.} Following the notation introduced by \citet{pasto13}, we label this early peak as Event A.
We note a moderately blue colour at the time of the Event A maximum,  $g-r \sim 0.10$ mag. 

\begin{table*}
\caption{\label{speclog} Log of the spectroscopic observations of \object{SN~2023ldh}. }
\centering
\begin{tabular}{ccccccc}
\hline\hline
Date (UT) & MJD & Phase &Instrumental configuration & Exptime (s) & Range (\AA) & Res.$^{(\dag)}$ (\AA) \\
\hline
2023-08-17   &  60173.95  & -63.05 & 10.4m GTC + OSIRIS + R1000B                & 3600       &   3600-7780  & 6.8     \\
2023-08-26   &  60182.90  & -54.10 & 10.4m GTC + OSIRIS + R1000B + R1000R       & 1200+1200  &   3600-10250 & 6.7+7.6 \\
2023-09-19   &  60206.88  & -30.12 & 3.56m TNG + Dolores + LRB                  & 3000       &   3500-7860  & 14.5    \\
2023-10-09   &  60226.74  & -10.26 & 1.82m Copernico + AFOSC + VPH7             &  624       &   3400-7280  & 14.5    \\
2023-10-10   &  60227.84  &  -9.16 & 1.82m Copernico + AFOSC + VPH7             & 2700       &   3400-7280  & 14.4    \\
2023-10-11   &  60228.77  &  -8.23 & 1.82m Copernico + AFOSC + VPH7             & 3600       &   3280-3280  & 14.8    \\
2023-10-12   &  60229.84  &  -7.16 & 10.4m GTC + OSIRIS + R2500V + R2500R       &  600+600   &   4415-7645  & 2.7+3.4 \\
2023-10-13   &  60230.85  &  -6.15 & 3.56m TNG + Dolores + LRR                  & 3600       &   5130-10270 & 11.1    \\
2023-10-14   &  60231.82  &  -5.18 & 2.56m NOT + ALFOSC + gm4                   & 1800       &   3400-9670  & 14.0    \\
2023-10-16   &  60233.84  &  -3.16 & 2.56m NOT + ALFOSC + gm4                   & 1200       &   3400-9670  & 13.8    \\
2023-10-17   &  60234.85  &  -2.15 & 1.82m Copernico + AFOSC + VPH7             & 1500       &   3400-7280  & 14.4    \\
2023-10-25   &  60242.81  &   5.81 & 10.4m GTC + OSIRIS + R1000B                &  400       &   3600-7780  &  6.8    \\
2023-11-03   &  60251.75  &  14.75 & 1.82m Copernico + AFOSC + VPH7             & 1800       &   3300-7270  & 14.4    \\
2023-11-07   &  60255.74  &  18.74 & 1.82m Copernico + AFOSC + VPH7             & 1800       &   3500-7270  & 14.5    \\
2023-11-08   &  60256.73  &  19.73 & 1.82m Copernico + AFOSC + VPH6             &  900       &   5000-9280  & 15.1    \\
2023-11-08   &  60256.74  &  19.75 & 1.82m Copernico + AFOSC + VPH6 (4" slit) &  900       &   5000-9280  & 37      \\
2023-11-14($\star$)   &  60262.72  &  25.72 & 1.82m Copernico + AFOSC + VPH7             & 1100       &   3400-7270  & 14.4    \\
2023-11-15   &  60263.20  &  26.20 & 1.82m Copernico + AFOSC + VPH7             & 2400       &   3400-7270  & 14.4    \\
2023-11-17   &  60265.19  &  28.19 & 1.82m Copernico + AFOSC + VPH6             & 2800       &   4990-9280  & 15.3    \\
2023-11-17   &  60265.75  &  28.75 & 1.82m Copernico + AFOSC + VPH7             & 2800       &   3300-7270  & 14.4    \\
2023-11-30   &  60277.26  &  40.26 & 2.56m NOT + ALFOSC + gm4                   & 2400       &   3400-9620  & 17.5    \\
2023-12-08   &  60286.26  &  49.26 & 2.56m NOT + ALFOSC + gm4                   & 2700       &   3400-9620  & 17.6    \\
2023-12-16   &  60294.26  &  57.26 & 2.56m NOT + ALFOSC + gm4                   & 3600       &   3400-9680  & 13.7    \\
2023-12-27   &  60305.25  &  68.25 & 2.56m NOT + ALFOSC + gm4                   & 3600       &   3400-9680  & 13.3    \\ 
2024-01-08($\star$)&60317.26& 80.26 & 2.56m NOT + ALFOSC + gm4                   & 3600       &   3400-9680  & 13.6    \\
2024-01-09   &  60318.24  &  81.24 & 2.56m NOT + ALFOSC + gm4                   & 3600       &   3400-9680  & 13.6    \\
2024-01-29   &  60338.21  & 101.21 & 2.56m NOT + ALFOSC + gm4                   & 5400       &   3400-9680  & 13.7    \\
2024-02-27   &  60367.20  & 130.20 & 10.4m GTC + OSIRIS + R1000R                & 2600       &   5050-10300 & 7.8     \\
2024-03-03   &  60372.37  & 135.37 & 6.5m MMT + BINOSPEC + 270/mm + LP3800& 3600       &   3830-9200  & 4.1     \\
2024-05-01   &  60431.57  & 194.57 & Keck-I + LRIS + gr.600/4000 + gr.400/8500 & 3600 & 3150-10200 & 3.4+5.8 \\
2024-05-04   &  60434.12  & 197.12 & 10.4m GTC + OSIRIS + R1000R                & 3600       &   5090-10300 & 7.5      \\   \hline
\end{tabular}
\tablefoot{The phases are computed from the epoch of the $r$-band maximum light of Event B (MJD = 60237.0 $\pm$ 0.4). The spectra marked with the ($\star$) symbol 
have poor signal-to-noise ratios, hence they are not shown in Fig. \ref{spectra}. $^{(\dag)}$ Computed from the FWHM of the [O~I] $\lambda$5577.3 and $\lambda$6300.3 night skylines.}
\end{table*}

Then, the light curves decline for about three weeks, until MJD~$= 60221.3 \pm 2.5$, to finally rise again, much more rapidly than before. \object{SN~2023ldh} reaches the $r$-band peak of the so-called Event B on MJD~$= 60237.0 \pm 0.4$ ($r = 15.74 \pm 0.02$ mag, i.e. $M_r = -18.53 \pm 0.23$ mag). 
At this second maximum, the colour is even bluer, $g-r \sim -0.22$ mag. 
As usually happens in SNe, the light-curve peak is reached earlier in the blue bands than in the redder bands. The epochs, the apparent and the absolute magnitudes at the maximum of Event B are reported in Table \ref{photpeak} for all filters. The maximum is followed by an initial rapid decline lasting approximately 40 days, with a decline rate of $\gamma_r = 5.2 \pm 0.1$ mag~(100~d)$^{-1}$ in the $r$-band light curve. The decline rate is much faster in the blue than in the red bands.\footnote{We estimate the following decline rates in the other bands: $\gamma_u = 8.4 \pm 0.3$ mag~(100~d)$^{-1}$, $\gamma_g = 6.5 \pm 0.1$ mag~(100~d)$^{-1}$, $\gamma_i = 4.9 \pm 0.1$ mag~(100~d)$^{-1}$, $\gamma_z = 4.3 \pm 0.1$ mag~(100~d)$^{-1}$, $\gamma_B = 7.3 \pm 0.2$ mag~(100~d)$^{-1}$, $\gamma_V = 5.7 \pm 0.1$ mag~(100~d)$^{-1}$.} 
From about +40 to +70~d, the light curves show a short-lasting plateau,
more evident in the red bands ($\gamma_r = 1.6 \pm 0.1$ mag~(100~d)$^{-1}$), followed by a rapid decline lasting about 3 weeks.
From phase $\sim$ 100 to 200~d, the decline rate in the $r$ band is $\gamma_r = 0.85 \pm 0.03$ mag~(100~d)$^{-1}$, \footnote{The decline rates in this phase in the other bands are  $\gamma_g = 1.07 \pm 0.04$  mag~(100~d)$^{-1}$ $\gamma_i = 0.87 \pm 0.05$ mag~(100~d)$^{-1}$, 
$\gamma_z = 0.80 \pm 0.04$  mag~(100~d)$^{-1}$, $\gamma_V = 1.02 \pm 0.05$ mag~(100~d)$^{-1}$.} which is surprisingly similar to that expected from the radioactive decay of $^{56}$Co into $^{56}$Fe (see discussion in Sect. \ref{Sect:LTR}). 

At phases later than $\sim$200~d, a moderate steepening is observed in the optical light curves. For a $^{56}$Co-powered SN light curve, this is usually interpreted as a consequence of an enhanced $\gamma$-rays escape with the expansion of the ejecta. However, while an incomplete trapping produces a light curve with a faster decline than the one expected from the $^{56}$Co decay, its slope is not expected to change over time.
Alternatively, we can invoke an attenuated contribution of the ejecta-CSM interaction at late times as the shock front reaches the outermost CSM regions. The increased rate of the luminosity decline may also be interpreted as evidence of the 'snow-plow' phase described by \citep{svi12}, which may occur when the mass gathered by the forward shock is comparable to the mass of the ejecta. Under this regime, we expect a gradual change in the slope of the bolometric light curve proportional to $t^{-1.5}$ \citep[see, also, discussion in][]{mor13}. Finally, an increased decline rate at late phases can also be interpreted as evidence of dust formation, although infrared observations are not available to support this scenario.

\section{Spectroscopy} \label{sect:spec}

 Our spectra span the evolution of the transient from about 2 months before to about $200$~d after the Event B maximum. Basic information on the spectra is reported in Table \ref{speclog}.
 The NOT/ALFOSC, Copernico/AFOSC and GTC spectra were reduced using the {\sc FOSCGUI} pipeline.\footnote{{\sc FOSCGUI} is a graphical user interface aimed at extracting SN spectroscopy and photometry obtained with FOSC-like instruments. It was developed by E. Cappellaro. A package description can be found at \url{sngroup.oapd.inaf.it/foscgui.html}.}
The LRIS spectrum was reduced using the {\sc IDL LPipe} package \citep{per19}, while spectra obtained with other instruments were reduced using standard routines in the {\sc IRAF} environment. Regardless of the package used, the reduction steps for all instruments included bias subtraction, flat-field correction, optimal extraction of the 1-D spectrum, wavelength and flux calibrations, and telluric band corrections. The spectral fluxing was then checked with coeval photometry and, when necessary, a constant correction factor was applied. 

   \begin{figure*}
   \centering
   \includegraphics[width=15.8cm]{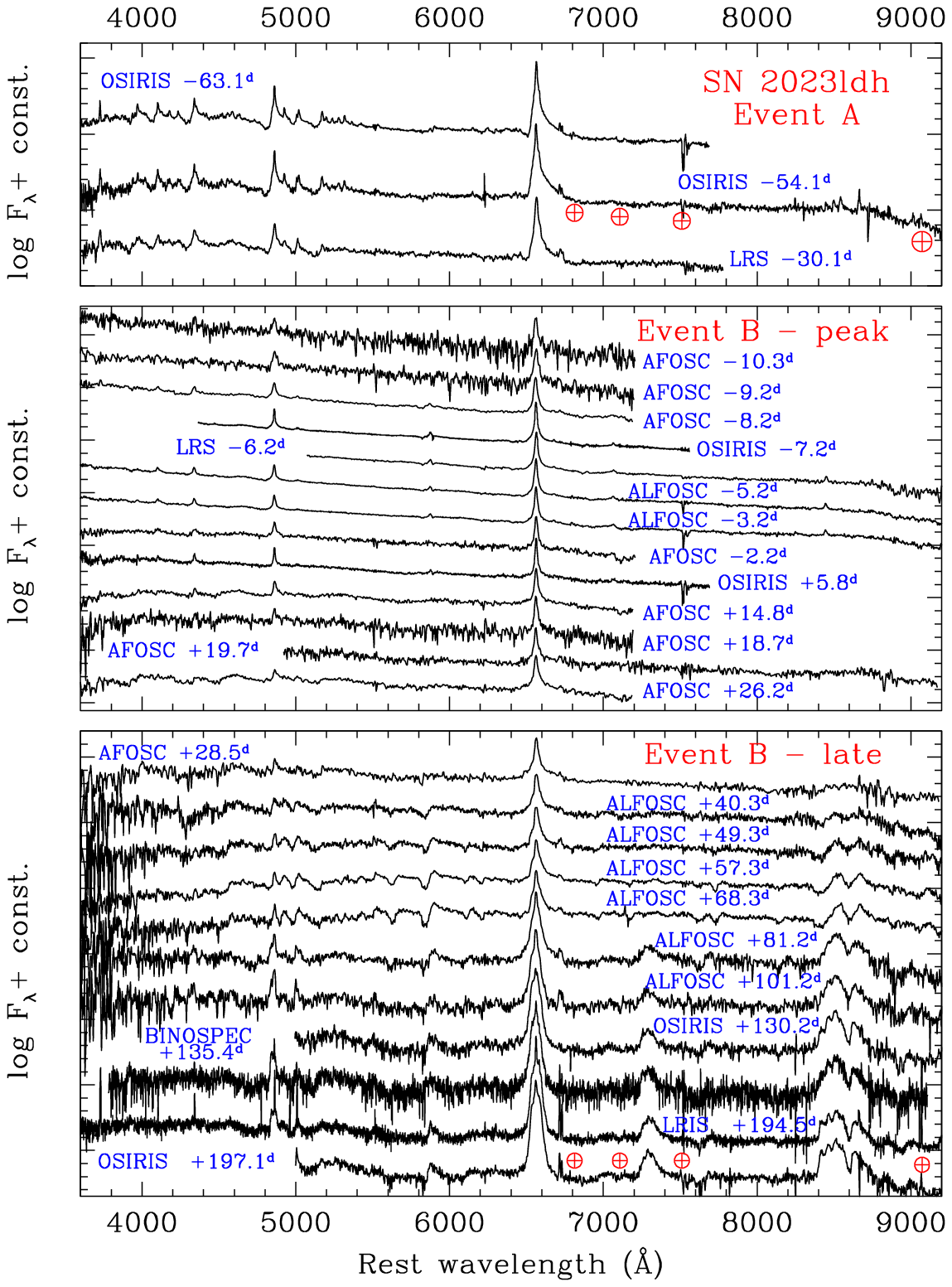}
      \caption{Sequence of optical spectra of \object{SN~2023ldh} obtained during Event A (top panel), around the maximum of Event B (middle panel) and at phases later than about four weeks from the Event B peak (bottom panel). The spectra are corrected for the redshift (z = 0.011321) of the  nearby source known as \object{SDSS J150912.08+523154.3}.}
         \label{spectra}
   \end{figure*}

\subsection{Line of sight reddening} \label{sect:reddening}

In order to estimate the total reddening in the direction of \object{SN~2023ldh} introduced in Sect. \ref{sect:intro},
we use the highest resolution GTC spectrum of the SN obtained $-7$~d from the maximum light (see Table \ref{speclog}) to estimate the reddening contribution of the host galaxy through the relation between the equivalent width (EW) of the narrow interstellar Na~I 5889, 5895 \AA~doublet (hereafter Na~ID) and the colour excess due to line-of-sight dust \citep{poz12}.\footnote{But see \citep{byr23} for a discussion on the limits of this method, at least when applied to another class of ejecta-CSM interacting transients.} Following the notation of \citet{poz12}, we measure $EW(D1)  = 0.40\pm0.03$, $EW(D2)  = 0.64\pm0.04$, and $EW(D1+D2)  = 1.13 \pm 0.04$ \AA. Using their equations 7 to 9, we obtain a weighted average colour excess for the host galaxy of $E(B-V)_{host} = 0.233 \pm 0.050$ mag, which has to be scaled by a factor of 0.86, as prescribed by \citet{das25}. Taking into account the Milky Way reddening towards \object{SN~2023ldh}, we obtain $E(B-V)_{tot}= 0.216 \pm 0.044$ mag, which is the value used for the total dust extinction in the direction of \object{SN~2023ldh}. 

\subsection{Spectral evolution} \label{sect:specevol}

The spectra of \object{SN~2023ldh} were analysed after applying redshift and reddening corrections. The adopted redshift near the SN location is z = 0.011321 (see Sect. \ref{sect:intro}). In order to accurately describe the spectroscopic evolution of \object{SN~2023ldh}, we identify five main evolutionary phases.

\begin{enumerate} 

\item Event A -- We collected spectra at three epochs during Event A, all of them showing a moderately hot continuum, with a black-body temperature of $T_{bb} \sim 9200$~K. The spectra show prominent lines of the Balmer series along with numerous P~Cygni lines of Fe~II. Na~ID and the near-infrared (NIR) Ca~II triplet are also clearly identified. Other metal lines are less conspicuous (e.g., Sc~II). The H lines show a composite profile: a narrower component peaking at the rest wavelength of the transition and having a full width at half-maximum (FWHM) \footnote{The FWHM values reported in this paper are corrected for spectral resolution broadening.} velocity ($v_{FWHM}$) of about 400 km s$^{-1}$, an intermediate-width component (slightly redshifted from the rest wavelength) with $v_{FWHM} \sim$ 2000-2500  km s$^{-1}$, and a broad component ($v_{FWHM} \sim 4500$ km s$^{-1}$) with a peak redshifted by more than 2000 km s$^{-1}$. We caution, however, that emission lines from background sources may affect the observed line profiles in these low-resolution spectra.
The H lines also show a multi-component absorption profile, which is well discernible in the first GTC/OSIRIS+ spectrum (at phase $-63.1$~d). In particular, in this spectrum H$\alpha$ shows a broad absorption with the minimum blue-shifted by $3760\pm300$ km~s$^{-1}$ and a blue wing extending up to over 4650 km~s$^{-1}$, plus a slower absorption component whose minimum is blueshifted by $2590\pm80$ km~s$^{-1}$. The profile of H$\beta$ is similar to that of H$\alpha$, although the measurements are complicated by line blending with P~Cygni Fe~II features.

   \begin{figure*}
   \centering
   \includegraphics[width=16.2cm]{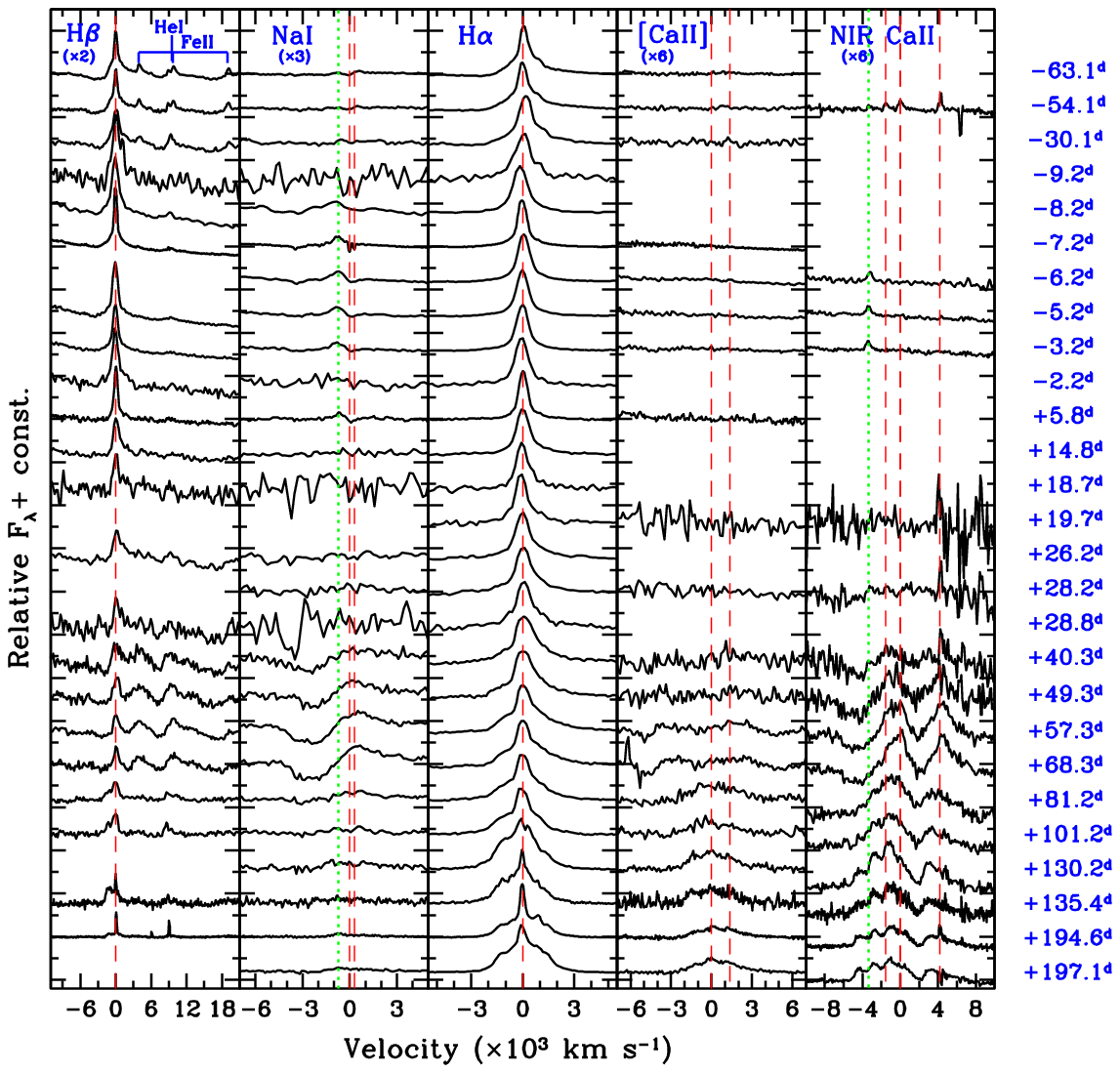}
      \caption{Evolution of the profiles of individual spectral lines. From the left to the right: H$\beta$, Na~ID (the green dotted line marks the position of He~I $\lambda$5876), H$\alpha$, [Ca~II] $\lambda\lambda$7291,7323, and the Ca~II NIR triplet (with the green dotted line marking O~I $\lambda$8446.5).}         \label{spectra_lineprof}
   \end{figure*}

\item Rise to Event B maximum -- After the Event~B onset, on MJD~$= 60221.3 \pm 2.5$, the continuum temperature rises and reaches $T_{bb} \sim$ 16000-17000 K a few days before the light-curve maximum, then gradually declines in the post-maximum evolution. We note that the spectral energy distribution (SED) peaks in the ultra-violet (UV) domain at these early epochs. Since this region is not covered by our optical spectra, the temperature estimates at the early phases obtained via blackbody fits to the spectral continuum are affected by large errors (typically $\pm$~2000-3000 K).
During the rise to the Event B peak (which lasts $\sim$15 days) and up to about three weeks past maximum, the SN spectrum shows very little evolution. It is characterized by a blue continuum with superposed prominent Balmer lines showing a dominant emission component. The line profile is best fitted with a Lorentzian function whose emission core is at the rest wavelength of the transition. The FWHM velocity of the H lines is $650 \pm 30$ km s$^{-1}$, as inferred from the best-resolution ($\sim150$ km s$^{-1}$) GTC/OSIRIS+ spectrum obtained $-7.2$~d from the maximum light. A residual broad P~Cygni absorption is detected at the positions of H$\beta$ and H$\gamma$ with a minimum blueshift of about 4000 km s$^{-1}$ and a blue wing that extends to about 8000-10000  km s$^{-1}$.
We also detect He~I lines ($\lambda$4471.5, $\lambda$4921.9, $\lambda$5015.7, $\lambda$5875.6, $\lambda$6678.1, $\lambda$7065.2), revealing a narrow emission component superposed on a broad P~Cygni base. While the FWHM velocity of the emission component is comparable to that of the narrow Balmer lines, He~I $\lambda$5875.6 shows a clear P Cygni profile, with a minimum at $-2780$~km s$^{-1}$ from the rest frame, and a blue-shifted wing extending up to about 5000~km s$^{-1}$. 
Another emission feature clearly detected in the low-resolution spectra from $-6.2$~d to $-3.2$~d from maximum is O~I $\lambda$8446.5, with $v_{FWHM} \sim$ 400-450~km~s$^{-1}$. 
Spectra showing a prominent O~I $\lambda$8446.5 emission line, while O~I $\lambda\lambda$7772-7775 remains below the detection threshold, are quite frequent in SNe IIn, such as \object{SN~1995G} \citep{pasto02} and  \object{SN~2010jl} \citep{fra14}. This feature is likely indicative of Ly$\beta$ pumping \citep[see][for a detailed discussion on the Bowen fluorescence mechanism]{val25}.
Finally, a broad emission bump is observed at 4600-4700~\AA, possibly due to a blend of C~III, N~III and He~II lines.

\item Early post-maximum decline (phases from 0~d to +30~d) -- At this phase, the spectrum shows a modest evolution, with a continuum becoming progressively redder.
The He~I lines become weaker, the absorption components of the Balmer lines disappear, and now the narrow H lines seem to sit on a broader base, with $v_{FWHM} \sim$ 7000-8000~km~s$^{-1}$. Unfortunately, the low signal-to-noise (S/N) ratio of some spectra and their modest resolution do not allow for a secure characterization of these line profiles, and the presence of electron-scattering wings cannot be ruled out. Some undulations appear in the bluest spectral region  (below $\sim$4800~\AA) about two weeks after maximum, becoming more prominent in later spectra. The features are likely due to the emerging contribution of broad Fe~II lines. 

\item Late post-maximum decline (phases from +40 to +70~d) -- This temporal window corresponds to the plateau in the light curve of \object{SN~2023ldh} shown in Fig. \ref{lightcurve}. In this phase, the spectral continuum becomes much redder and the undulations at the blue wavelengths, mostly due to blends of Fe~II lines, are now more evident. A good S/N spectrum was taken at +57.3~d, and it shows an evident flux deficit below $\sim$4600\AA, possibly due to the blanketing of metal lines. Along with the Balmer lines with a prominent emission component, broad P~Cygni features of Fe~II, Sc~II and Ba~II are now visible. The position of the blue-shifted minimum of the metal lines constrains the velocity at the photosphere in this phase. The photospheric velocity was measured for the relatively unblended Fe~II $\lambda$4923.9 (triplet 42) line. It declines from $2230 \pm 790$~km s$^{-1}$ at +40.3~d, to $2040 \pm 390$~km s$^{-1}$ at +49.3~d, to $1870 \pm 260$~km s$^{-1}$ at +57.3~d, and to $1680 \pm 230$~km s$^{-1}$ at +68.3~d. A similar evolution of the line velocity is also measured for the Sc~II  $\lambda$6247 (multiplet 28) feature, the Ba~II  $\lambda$6141.7 line (multiplet 2), and  O~I $\lambda\lambda$7772-7775. Noticeably, the possible detection in this phase of O~I $\lambda$8446.5 along with O~I $\lambda\lambda$7772-7775 would be consistent with the expectations for H recombination.
Other metal lines unequivocally detected in this spectrum are Na~ID (with a velocity systematically higher by 20-25$\%$ than other metal lines), and the Ca~II NIR triplet which becomes the most prominent spectral feature after H$\alpha$. H$\alpha$ has an evident asymmetric emission profile, still dominated by a narrow emission component ($v_{FWHM} \sim 500-600$ km s$^{-1}$) centered at the rest wavelength, an intermediate-width emission component (with $v_{FWHM} \sim 3000 $ km s$^{-1}$), and a slightly red-shifted  broader component, with a FWHM velocity fading from about 7600 km s$^{-1}$ at +40.3~d to $v_{FWHM} \sim 5300$ km s$^{-1}$ at +68.3~d.
The spectra obtained during the plateau globally resemble those of low-luminosity Type IIP SNe \citep[e.g.,][]{spiro14}, except for the still dominant Balmer line emission components. The overall spectral evolution of \object{SN~2023ldh} during Event B may also resemble those observed in some peculiar type II events such as \object{SN~2016bkv} \citep{nak18,hos18} and \object{SN~2018zd} \citep{zha20,hir21,cal21}, although their plateaus were longer lasting.

\item Nebular phase (phase $>$ +80~d) -- Two spectra of \object{SN~2023ldh} were obtained during the steep post-plateau luminosity decline (at +80.3~d and +81.2~d). They still show some P~Cygni features as in the previous phase, but now [Ca~II] $\lambda\lambda$7291-7323, a classical nebular feature observed in core-collapse SN spectra, starts to appear.
The following spectra, obtained during the late-time photometric flattening, become fully nebular, with P~Cygni features progressively vanishing, and both [Ca~II] (with a total FWHM velocity of 3500-3800 km s$^{-1}$) and the Ca~II NIR triplet becoming more prominent. We also note that O~I $\lambda$8446.5 now contributes to the broad blend with the  Ca~II NIR triplet. H$\alpha$ remains the most prominent spectral feature, but now the narrow line component is weaker and heavily contaminated by unresolved lines of the nearby H~II region (H$\alpha$, [N~II] and, more marginally, [S~II]). The profile of H$\alpha$ in the spectra obtained from 100 to 200~d after the maximum is dominated by the intermediate-width component, which now shows a boxy profile with a FWHM velocity of about 2500 km s$^{-1}$, and whose edges extend up to $\sim1400$ km s$^{-1}$ from the rest wavelength. H$\beta$ shows a similar profile as H$\alpha$, while the Na ID feature (perhaps blended with He~I $\lambda$5875.6) is still visible. We also note in the best-resolution nebular spectra that weak narrow features are visible on top of the dominating broad Ca~II lines. Similarly, weak and marginally resolved features are also visible at the position of the [O~I] $\lambda\lambda$6300-6364 doublet, and are likely produced in the unshocked CSM. A persistent blue pseudo-continuum (below $\sim 5600$\AA) remains visible also in the latest spectra of \object{SN~2023ldh} and, as for other ejecta-CSM interacting SNe, it is usually interpreted as due to blends of Fe lines \citep[e.g.,][]{smi09,sal25}.

\end{enumerate} 

The evolution of the profiles of individual spectral lines is reported in Fig. \ref{spectra_lineprof}, which shows the strengthening of permitted and forbidden Ca~II features in our latest spectra, along with the emergence of the strongly asymmetric H$\alpha$ profile. We note that weak and narrow P~Cygni lines of the Ca~II NIR triplet were visible only during Event A, with a velocity of 400 km s$^{-1}$. At the onset of Event B, the Ca~II NIR triplet is not identified, which is consistent with the rapid rise of the continuum temperature; a narrow, marginally resolved O~I $\lambda$8446.5 line is instead seen in pure emission. This line becomes much weaker in the post-maximum spectra, to re-emerge again in those at phases $\gtrsim+100\,\rm{d}$.

Finally, a broad Na~I feature with a P~Cygni profile is detected only from about 40 to 70 days past maximum. Its minimum is blue-shifted by 2600 km s$^{-1}$ from the rest wavelength, a velocity consistent with that inferred for the broader component of H$\alpha$. 

\subsection{Metallicity of the SN environment}  \label{Sect:metallicity}

Using the good resolution MMT spectrum obtained on 2024 March 3, we estimate the metallicity at the SN location. \object{SN~2023ldh} is, in fact, in close proximity to a luminous H~II region. We measure the flux of the H~II emission lines, in particular H$\alpha$, H$\beta$, as well as the [O III] and [N II] doublets. According to \citet{mar13}, the above lines are good indicators of the local abundance of oxygen.

Following \citet{mar13}, we first determine the O3N2 parameter from the flux ratios of four strong spectral lines.

   \begin{equation}
O3N2 = log \left(\frac{[O~III]_{\lambda5007}}{H\beta} \times \frac{H\alpha}{[N~ II]_{\lambda6583}}\right).
   \end{equation}

   \begin{figure}
   \centering
   \includegraphics[width=9.0cm]{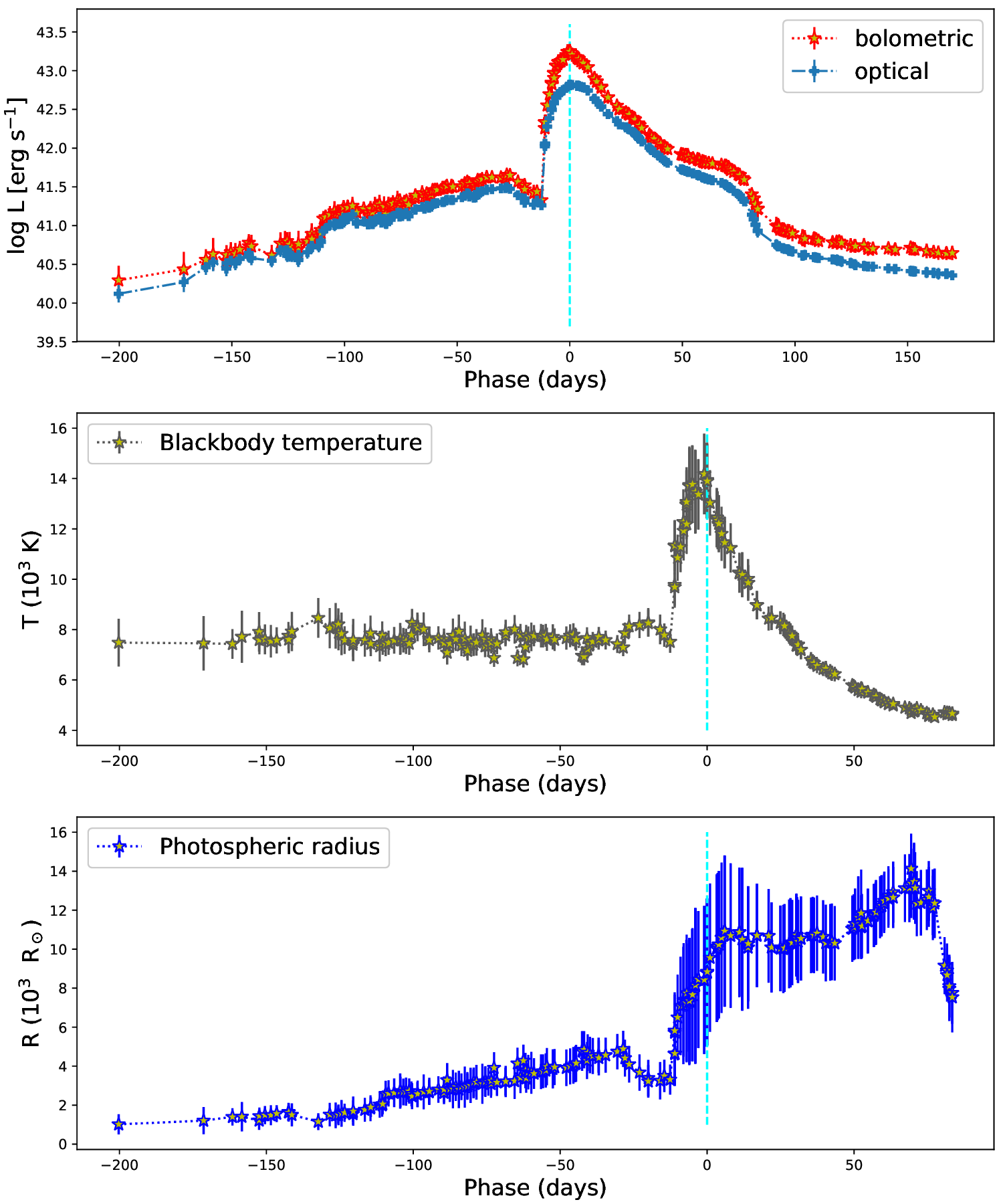}
      \caption{Bolometric and quasi-bolometric (optical domain only) light curves of SN~2023ldh (top panel); black-body temperature evolution (middle panel) 
and evolution of the photospheric radius 
(bottom panel). The phases are from the bolometric light-curve maximum (MJD~$= 60234.8 \pm 1.0$), marked with a vertical line.}
         \label{LTR}
   \end{figure}

\noindent The O3N2 parameter is a good proxy for the oxygen abundance when $-1.1 <$~O3N2~$< 1.7$, which is the case of the H~II region in the proximity of the SN (O3N2 $\approx 0.26$). The oxygen abundance near the SN location is obtained through equation 2 of \citet{mar13}:
 
   \begin{equation}
12 + log (O/H) = 8.533 - 0.214 \times O3N2 = 8.48 \pm 0.03.
   \end{equation}

An oxygen abundance of 8.48 dex implies a slightly sub-solar metallicity for the H~II region closest to the location of \object{SN~2023ldh}, which is consistent with the one inferred from the statistical approach of \citet{pyl04} based on the macroscopic properties of the host galaxy (morphological type, absolute magnitude, inclination), and computed at the SN location (12 + log (O/H) $\approx$ 8.47 dex). However, we note that other calibrations of the O3N2 diagnostic \citep[e.g.,][]{cur17} would point towards higher values for the oxygen abundance. 

\section{Bolometric light curve and temperature/radius evolution} \label{Sect:LTR}

Although our photometric monitoring of \object{SN~2023ldh} is limited to the optical domain, we can provide a guess of the evolution of the bolometric luminosity by performing a black-body fit to the SED. The fluxes at individual epochs are inferred from the available $u$ to $z$ magnitudes after correction for interstellar extinction. When observations are not available in individual filters, the fluxes are obtained through an interpolation from the available photometry at adjacent epochs, or by extrapolating the fluxes assuming a constant colour evolution. The integration of the best-fit black-body at a certain epoch over the entire wavelength range allows us to estimate the bolometric luminosity. We also obtain a pseudo-bolometric light curve that includes only the contribution of the observed optical bands. For each epoch, we measure the flux contributions of the different filters using the trapezoidal rule, assuming a negligible flux contribution at the integration extremes.

The resulting ``optical'' quasi-bolometric and bolometric light curves are shown in Fig. \ref{LTR} (top panel). The evolution of the black-body temperature ($T_{bb}$), as inferred from the SED fits, is shown in the middle panel. Finally, given the luminosity and temperature information, we can estimate the evolution of the radius at the photosphere ($R_{ph}$; Fig. \ref{LTR}, bottom panel).

   \begin{figure*}
   \centering
   \includegraphics[width=13.9cm,angle=270]{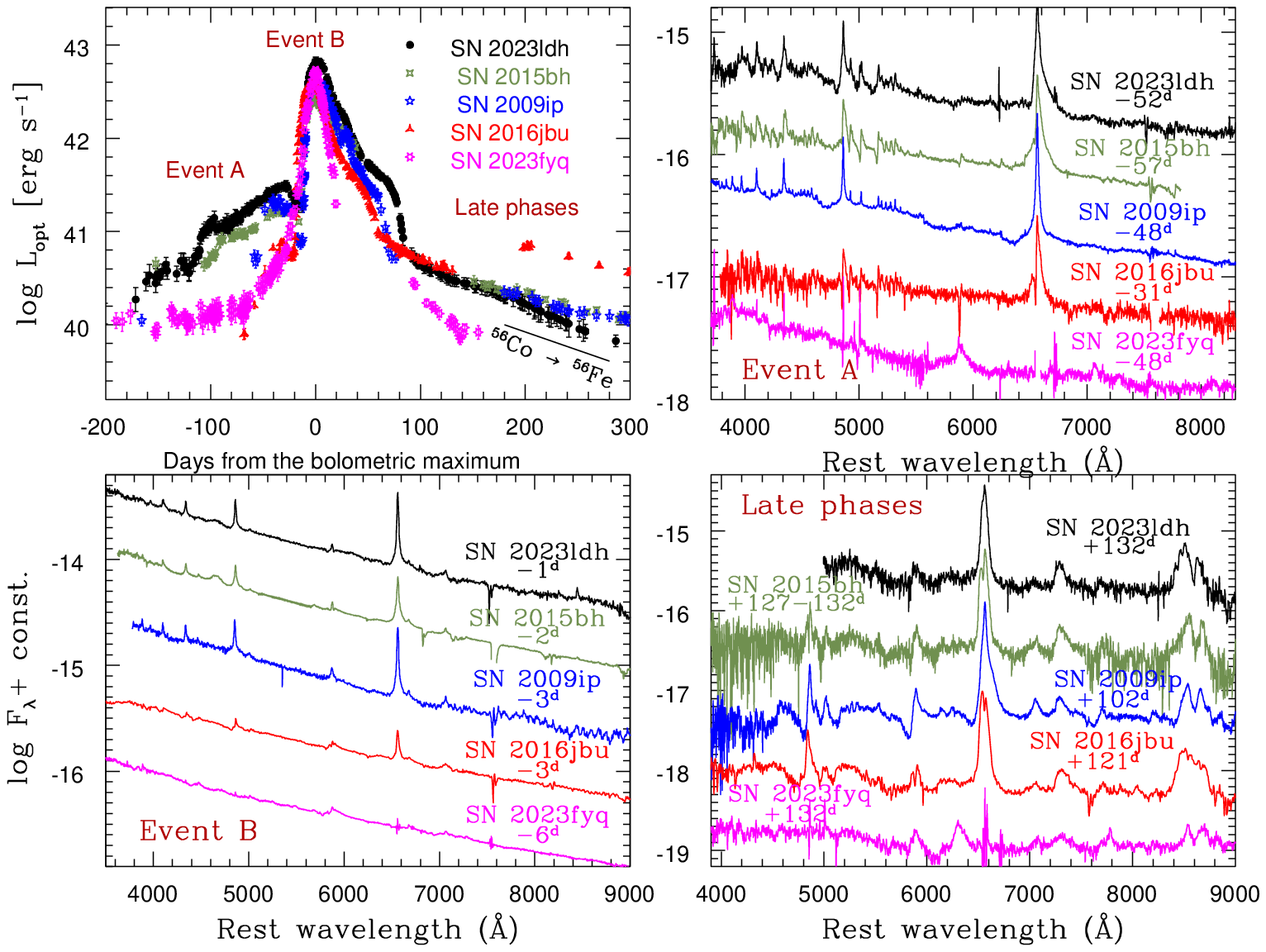}
      \caption{Top-left panel: Comparison of the pseudo-bolometric light curve of \object{SN~2023ldh}, obtained integrating the broad-band fluxes in the optical domain only, with those of the type IIn SNe 2009ip, 2015bh, 2016jbu and the type Ibn SN~2023fyq \citep{bre24,dong24}. Other panels: comparison of the optical spectra of SNe~2023ldh, 2009ip, 2015bh, 2016jbu and 2023fyq at similar phases: from about $-60$ to $-30$~d (during Event A; top-right panel), a few days before the maximum of Event B (bottom-left panel); at late phases (100-130~d after the bolometric peak; bottom-right panel). The data of SN~2009ip are  from \protect\citet{pasto13}, \protect\citet{fra13,fra15}, \protect\citet{mar14}, \protect\citet{pri13}, \protect\citet{gra14}; those of SN~2015bh are from \citet{nancy16}, \citet{gor16}, \citet{ofe16}, \citet{tho17}; and those of SN~2016jbu from \citet{bre22a}. The spectra of SN~2023fyq shown here are those of \protect\citet{bre24} and \protect\citet{dong24}, after a careful subtraction of nearby contaminating sources performed using a spectrum of the host environment obtained at the NOT telescope on 2023 December 25, when the SN was no longer visible.}
      \label{comparison}
   \end{figure*}

The bolometric light curve shows a global but non-monotonic rise lasting about six months before reaching a local maximum at $-30$~d, which can be considered the peak
of Event A, at a bolometric luminosity of $L^A_{bol}  \sim 4.5 \times 10^{41}$ erg s$^{-1}$. However, some luminosity fluctuations are clearly visible in this phase, with the strongest being at phase $-100$~d.  The maximum of Event A is followed by a significant luminosity decline lasting about 2 weeks. At this minimum, the bolometric luminosity is a factor two lower than at the time of the Event A maximum. The subsequent evolution shows a rapid luminosity rise during Event B, with a bolometric maximum reached on MJD~$= 60234.8 \pm 1.0$ (with $L^{B}_{bol} = 1.78 (\pm 0.43) \times 10^{43}$ erg s$^{-1}$). In this phase, likely dominated by the UV emission (as indicated by the hot spectral continuum, see Sect. \ref{sect:spec}), the optical domain contributes to the total luminosity by less than 40 per cent. After maximum, the bolometric light curve monotonically declines until day +50. From about +50 to +75~d, the light curve shows a sort of plateau, at the end of which it fades very rapidly for about two weeks, to then flatten again from phase +90~d onward. However, bolometric luminosity estimates at phases later than $\sim$ +90~d should be taken with caution as the SED largely deviates from a black-body.

As mentioned in Sect. \ref{sect:lc}, the luminosity decline rates at phases later than about three months past maximum are consistent with the expectations for a $^{56}$Co decay-powered light curve. Assuming that the late luminosity is solely powered by radioactive decays, the amount of $^{56}$Ni synthesized can be inferred from a comparison with the bolometric luminosity of \object{SN~1987A} \citep{cat88} at the same phase after the explosion. With this method, we obtain a $^{56}$Ni mass of about 0.015 M$_\odot$ for \object{SN~2023ldh}. However, we caution that ejecta-CSM interaction signatures are clearly visible in the SN spectra (see Sect. \ref{sect:specevol}), hence we should consider this value as a conservative upper limit for the amount of $^{56}$Ni ejected by \object{SN~2023ldh}.

The black-body fit to the SED provides direct information on the evolution of the temperature. During the pre-SN phases, $T_{bb}$ remains in the 7000 to 8000~K range, although some fluctuations can be noticed. $T_{bb}$ reaches about 8300~K at the time of the Event A peak\footnote{We note that this temperature is lower by almost 1000~K from that inferred by fitting the spectral continuum (see Sect. \ref{sect:specevol}).}, then it slightly declines to about 7500~K soon before the onset of the Event B brightening. The $T_{bb}$ of \object{SN~2023ldh} reaches $\sim$14000~K at the time of the Event B bolometric luminosity maximum. Later, the temperature monotonically declines to reach $\sim$5000~K about two months after the maximum. We remark that at phases later than $\sim$+80 d, the spectrum is largely dominated by emission lines, which makes $T_{bb}$ measurements no longer meaningful.

The photospheric radius $R_{ph}$ increases very slowly during Event A from $\sim$ 1000 to 5000 R$_\odot$ at $-30$~d, then it decreases slightly for about two weeks. At the onset of \object{SN~2023ldh} Event B, $R_{ph}$ increases very rapidly to the bolometric peak and remains nearly constant from phase $\sim$0 to $+50$~d, at about 10500 R$_\odot$. During the plateau in the bolometric light curve, the radius rises again until $R_{ph}$ $\sim$ 14000  R$_\odot$ (at about $+70$~d), while the photosphere rapidly recedes at later phases.

\section{Comparison with other SN~2009ip-like events} \label{Sect:cfr}

In Fig. \ref{comparison} we compare the observed properties of \object{SN~2023ldh} with those of photometrically similar SNe IIn for which Event A was well sampled both in photometry and spectroscopy: \object{SN~2009ip}, \object{SN~2015bh} \citep{nancy16,gor16,ofe16,tho17} and \object{SN~2016jbu} \citep{bre22a,bre22b}. The 
type Ibn \object{SN~2023fyq} \citep{bre24,dong24} is also considered as a comparison object, although spectroscopically different and with a lower-luminosity Event A than other objects of the sample. 
The ``optical'' pseudo-bolometric light curves of the five objects are shown in the top-left panel, while the other three panels show their optical spectra at three reference phases: during the rise to the Event A maximum (phase $\sim -55$~d; top right), in the proximity of the Event B peak (phases $\sim -6<$ to $-1$~d; bottom-left), and at late phases ($> +100$~d; bottom-right). 

\object{SN~2023ldh} is marginally brighter than other \object{SN~2009ip}-like events at all phases, with a slightly longer duration of both Events A and B (Fig. \ref{comparison}, top-left panel). The late-time luminosity is virtually identical for all objects at the same epochs, with the exception of \object{SN~2023fyq} that is significantly fainter. All SNe are characterized by a similar decline rate after the Event B maximum, consistent with that expected for light curves powered by the radioactive decay of $^{56}$Co into $^{56}$Fe. If these bolometric light curves are powered
by radioactive decays, the amount of $^{56}$Ni synthesized in these explosions should necessarily be similar for all \object{SN~2009ip}-like objects of our sample (see, however, discussion in Sect. \ref{Sect:LTR}).
The observational analogies for the objects of our sample after years-long phases of erratic variability (Fig. \ref{comparison}) are somewhat surprising, and may imply similarity in the progenitor systems, the CSM geometry and the explosion mechanisms.

The spectra of the objects shown in Fig.\ref{comparison} display some differences in the continuum and the line profiles during Event A. In contrast, they appear to be quite similar during Event B, although with some subtle differences, in particular in the strengths and widths of the He~I lines. The He~I features are quite prominent in \object{SN~2009ip}, relatively weak in \object{SN~2015bh}, and almost negligible in \object{SN~2023ldh}. This suggests either a difference in the ionisation state of the line-forming regions of the three objects or some intrinsic differences in the abundance of He-rich material.
Recently, \citet{bre24} and \citet{dong24}  reported observations of the type Ibn \object{SN~2023fyq}. Despite the different spectroscopic classification, this SN~Ibn shows remarkable similarity in the photometric properties with our \object{SN~2009ip}-like sample. In particular, its light curve has a long-lasting pre-SN eruptive phase reminiscent of an Event~A, followed by a more luminous Event~B.

Other ejecta-CSM interacting SNe share similarities with the above objects, including several type IIn SNe\footnote{The group of SN~2009ip-like type IIn SNe includes also \object{SN~2010mc} \citep{ofe13}, \object{iPTF13z} \citep{nyh17}, SN~2013gc \citep{reg19}, \object{SNhunt151} \citep{nancy18}, \object{LSQ13zm} \citep{tar16}, SN~2016cvk \citep{mat25}, \object{SN~2016bdu} \citep{pasto18}, \object{SN~2018cnf} \citep{pasto19b}, SN 2019zrk \citep{fra22}, SN~2021qqp \citep{hir24} and the 2024 event following SN~2022mop \citep{bre25}.} and two transitional type IIn/Ibn events, \object{SN~2021foa} \citep{reg22,far24,gan24} and \object{SN~2022pda} (Cai et al., in prep.). However, in all these cases, the available spectroscopic and photometric data do not completely cover the crucial phases of the SN evolution illustrated in Fig. \ref{comparison}.

\section{Eruption or supernova explosion?} \label{Sect:discussion}

Long-lasting eruptive phases accompanied by a global trend of slow luminosity rise are typical of the so-called luminous red novae \citep[LRNe, e.g.,][and references therein]{pasto23}, which are thought to result from merging events of non-degenerate stars. For example, \object{SN~2009ip}-like events show rapid photometric variability followed by a slow brightening, somewhat reminiscent of the pre-outburst phase of LRNe \citep[e.g., V1309~Sco;][]{tyl11,sok13}. 

\citet{sch20} suggested that some SNe IIn are the outcome of the coalescence of a compact object, a black hole (BH) or a neutron star (NS), with the core of a massive companion, while the system is embedded in a H-rich common envelope.  
\citet{bre25} presented a study on \object{SN~2022mop}, a transient source characterized by two spatially coincident outbursts, both with duration and luminosity consistent with a SN explosion. 
\citet{bre25} proposed that the first event in 2022 was a genuine stripped-envelope SN in a binary system, whose remnant - likely a NS in an unstable eccentric orbit - started to interact with the outer layers of the H-rich secondary producing the photometric fluctuations observed in 2023. The interaction within a common envelope led the orbital period to gradually decrease, and the NS to merge into the secondary producing a \object{SN~2009ip}-like event in 2024.

A somewhat similar scenario is proposed by \citet{tsu24} for the type Ibn \object{SN~2023fyq}, whose properties favour a binary system with a He donor star transferring mass to a NS (or BH) companion. The super-Eddington mass transfer triggers a circum-binary outflow 
from the Lagrangian point L$_2$ and a fast wind from the accretion disk of the degenerate companion, which power the slow-rising light curve of \object{SN~2023fyq} in the years preceding the main event. The loss of angular momentum leads the two stars to merge, producing an outburst with the properties of an SN Ibn, without necessarily a terminal core collapse. 
According to \citet{tsu24}, the above model may only explain transients with smooth brightenings. Hence, while the scenario is applicable to the evolution of SNe IIn such as \object{SN~2021qqp} \citep[][]{hir24} and SNhunt151 \citep{nancy18}, it cannot comfortably explain \object{SN~2009ip}-like events characterized by former erratic, fast-evolving variability \citep[but, see][]{bre25}.

Whether the rapid luminosity fluctuations of \object{SN~2023ldh} from $\sim 4.5$ to $\sim1$ years before Event A are indicative of binary interaction or erratic variability similar to LBVs\footnote{An example is V37 in NGC~2403 \protect\citep{weis05,mau06,hum17}.} is unclear. Well-cadenced data with high S/N are required to eventually reveal periodicity in the archival observation. However, we note that some periodicity in the mass-loss events suggesting binary interaction was revealed by the bumps in the light curve of Event B in \object{SN~2009ip} \citep{kas13,mar15}.

Eruptive wave-driven outbursts \citep[e.g.,][]{qua12,mor14,leu20,chu22} or steady wind mass-loss resulting from a continuous energy deposition at the base of the stellar envelope in progenitors exceeding the Eddington's luminosity are still plausible explanations for the pre-SN evolution of \object{SN~2023ldh} and similar events \citep{mat22}. Mass loss generates an optically thick CSM engulfing the progenitor. The subsequent stellar core-collapse triggers high-velocity ejecta which collide with the CSM, sustaining the Event~B light curve. However, this scenario cannot  reproduce the frequent short-duration flares at $M_r \sim -14$ observed before Event~A.

Another scenario was proposed for \object{SN~2009ip}-like objects, according to which the entire light curve is explained through a non-terminal eruptive phase of a very massive star. In particular, the CSM is produced after a long-lasting (years) period of erratic variability, during which the star loses a significant amount of envelope mass. Later, the star experiences a giant eruption (Event A) and the interaction between the material ejected in Event A with the previously gathered CSM powers Event B. In this case, the terminal SN explosion is not required for reproducing the SN observables. However, only very massive stars can release enough mass and radiate such high energy for a long time without the need to invoke a terminal explosion \citep[e.g., in a pulsational pair-instability event;][]{woo17}.

Ultimately, we need to understand the fate of the progenitor (or the progenitor system) of \object{SN~2023ldh}. The study of putative survivors of former \object{SN~2009ip}-like transients gives somewhat contradictory results. While in most cases no residual source is visible at the SN location (or the source is orders of magnitudes fainter than the original quiescent progenitor) in deep images taken with space telescopes \citep{jen22,bre22c,smi22}, in at least one case (\object{SN~2011fh}) a residual source remains visible a long time after the putative SN explosion \citep[][]{pes22,reg24}, hence the survival or the death of the progenitor stars of these interacting SNe still remains an open issue. As for similar objects, only observations of the explosion site years after discovery and in multiple domains will clarify if its progenitor has survived the \object{SN~2023ldh} explosion. If a source will still be detectable, a multi-band study is expected to provide information on the bolometric luminosity and the spectral type of the survivor, which is key to retrospectively constrain the physical scenario producing the observed chain of events culminating in  \object{SN~2023ldh}. However, the late-time decline rates compatible with the $^{56}$Co decay and the similar synthesized $^{56}$Ni masses inferred for \object{SN~2023ldh} and the comparison objects in Sect. \ref{Sect:cfr} would favour similar progenitors and terminal core-collapse explosions for most \object{SN~2009ip}-like objects.

The next-generation surveys, such as the Legacy Survey of Space and Time \citep{ham23} at the Vera Rubin Observatory, will provide a harvest of new SNe IIn with precursors. The deep photometry released by the survey complemented with ancillary programs at medium-size and large telescopes will provide key information on the progenitors and the nature of their variability, while the deep and spatially resolved imaging of space telescopes ({\it Hubble~Space~Telescope}, {\it James Webb Space Telescope}, and {\it Euclid}) will reveal the fate of the star after the main explosive event. In particular, very late-time observations from 0.2 to 15 $\mu$m will provide bolometric luminosity estimates for any accreting source or remaining star.

\section*{Data availability}
Table A.1 only available in electronic form at the CDS via anonymous ftp to cdsarc.u-strasbg.fr (130.79.128.5) or via http://cdsweb.u-strasbg.fr/cgi-bin/qcat?J/A+A/.

\begin{acknowledgements}

AP, AR, GV, NER, PO and IS acknowledge support from the PRIN-INAF 2022 ``Shedding light on the nature of gap transients: from the observations to the models''.
NER acknowledges the Spanish Ministerio de Ciencia e Innovaci\'on (MCIN) and the Agencia Estatal de Investigaci\'on (AEI) 10.13039/501100011033 under the program Unidad de Excelencia Mar\'ia de Maeztu CEX2020-001058-M.
YZC is supported by the National Natural Science Foundation of China (NSFC, Grant No. 12303054), the National Key Research and Development Program of China (Grant No. 2024YFA1611603), the Yunnan Fundamental Research Projects (Grant Nos. 202401AU070063, 202501AS070078), and the International Centre of Supernovae, Yunnan Key Laboratory (No. 202302AN360001).
LG acknowledges financial support from AGAUR, CSIC, MCIN and AEI 10.13039/501100011033 under projects PID2023-151307NB-I00, PIE 20215AT016, CEX2020-001058-M, ILINK23001, COOPB2304, and 2021-SGR-01270.
PC and TLK acknowledge support via the Research Council of Finland (grant 340613).
TK acknowledges support from the Research Council of Finland project 360274.
SS has received funding from the European Union’s Horizon 2022 research and innovation programme under the Marie Sk\l{}odowska-Curie grant agreement No 101105167 — FASTIDIoUS.
MDS is funded by the Independent Research Fund Denmark (IRFD, grant number  10.46540/2032-00022B) and by an Aarhus Univesity Research Foundation Nova project (AUFF-E-2023-9-28).
SV and the UC Davis time-domain research team acknowledge support by NSF grants AST-2407565CPG acknowledges financial support from the Secretary of Universities
and Research (Government of Catalonia) and by the Horizon 2020 Research
and Innovation Programme of the European Union under the Marie
Sk\l{}odowska-Curie and the Beatriu de Pin\'os 2021 BP 00168 programme,
from the Spanish Ministerio de Ciencia e Innovaci\'on (MCIN) and the
Agencia Estatal de Investigaci\'on (AEI) 10.13039/501100011033 under the
PID2023-151307NB-I00 SNNEXT project, from Centro Superior de
Investigaciones Cient\'ificas (CSIC) under the PIE project 20215AT016
and the program Unidad de Excelencia Mar\'ia de Maeztu CEX2020-001058-M,
and from the Departament de Recerca i Universitats de la Generalitat de
Catalunya through the 2021-SGR-01270 grant.

This work makes use of data from the Las Cumbres Observatory global telescope network. The LCO group is supported by NSF grants AST-1911151 and AST-1911225.\\ 

The data presented herein were obtained in part with ALFOSC, which is provided by the Instituto de Astrofisica de Andalucia (IAA) under a joint agreement with the University of Copenhagen and NOT.
This work is based on observations made with the Nordic Optical Telescope (NOT), owned in collaboration by the University of Turku and Aarhus University, and operated jointly by Aarhus University, 
the University of Turku and the University of Oslo, representing Denmark, Finland and Norway, the University of Iceland and Stockholm University at the Observatorio del Roque de los Muchachos, 
La Palma, Spain, of the Instituto de Astrofisica de Canarias; the 10.4\,m Gran Telescopio Canarias (GTC), installed in the Spanish Observatorio del Roque de los 
Muchachos of the Instituto de Astrof\'isica de Canarias, in the Island of La Palma;  the 2.0\,m Liverpool Telescope operated on the island of La Palma by Liverpool John Moores University 
at the Spanish Observatorio del Roque de los Muchachos of the Instituto de Astrof\'isica de Canarias with financial support from the UK Science and Technology Facilities Council; 
the 3.58\,m Italian Telescopio Nazionale Galileo (TNG) operated the island of La Palma by the Fundaci\'on Galileo Galilei of the Istituto Nazionale di Astrofisica (INAF) at the Spanish 
Observatorio del Roque de los Muchachos of the Instituto de Astrof\'isica de Canarias; the 1.82\,m Copernico and the 67/92\,cm Schmidt telescopes of INAF --- Osservatorio Astronomico di Padova, Asiago, Italy.
The observations reported here were obtained at the MMT Observatory, a joint facility of the Smithsonian Institution and the University of Arizona.
The LRIS data presented herein were obtained at the W. M. Keck Observatory from telescope time allocated to NASA through the agency's scientific partnership with the California 
Institute of Technology and the University of California. The Observatory was made possible by the generous financial support of the W. M. Keck Foundation.\\
This work has made use of data from the Asteroid Terrestrial-impact Last Alert System (ATLAS) project. ATLAS is primarily funded to search for near-Earth objects through NASA 
grants NN12AR55G, 80NSSC18K0284, and 80NSSC18K1575; byproducts of the NEO search include images and catalogs from the survey area. The ATLAS science products have been made 
possible through the contributions of the University of Hawaii Institute for Astronomy, the Queen's University Belfast, STScI, and the South
 African Astronomical Observatory, and The Millennium Institute of Astrophysics (MAS), Chile.\\
The Pan-STARRS1 Surveys (PS1) and the PS1 public science archive have been made possible through contributions by the Institute for Astronomy, the University of Hawaii, the Pan-STARRS Project Office, the Max-Planck Society and its participating institutes, the Max Planck Institute for Astronomy, Heidelberg and the Max Planck Institute for Extraterrestrial Physics, Garching, The Johns Hopkins University, Durham University, the University of Edinburgh, the Queen's University Belfast, the Harvard-Smithsonian Center for Astrophysics, the Las Cumbres Observatory Global Telescope Network Incorporated, the National Central University of Taiwan, STScI, NASA under grant NNX08AR22G issued through the Planetary 
Science Division of the NASA Science Mission Directorate, NSF grant AST-1238877, the University of Maryland, Eotvos Lorand University (ELTE), the Los Alamos National Laboratory, and the Gordon and Betty Moore Foundation.\\
This publication is partially based on observations obtained with the Samuel Oschin 48-inch Telescope at the Palomar Observatory as part of the Zwicky Transient Facility (ZTF) project. 
ZTF is supported by the National Science Foundation under Grants No. AST-1440341 and AST-2034437 and a collaboration including current partners Caltech, IPAC, the Oskar Klein Center at Stockholm University, the University of Maryland, University of California, Berkeley , the University of Wisconsin at Milwaukee, University of Warwick, Ruhr University, Cornell University, Northwestern University and Drexel University. Operations are conducted by COO, IPAC, and UW.\\
This research has made use of the NASA/IPAC Extragalactic Database (NED) which is operated by the Jet Propulsion Laboratory, California Institute of Technology, under contract with NASA. This publication used data products from the Two Micron All-Sky Survey, which is a joint project of the University of Massachusetts and the Infrared Processing and Analysis Center/California Institute of Technology, funded by NASA and the NSF. \\
The authors wish to recognise and acknowledge the very significant cultural role and reverence that the summit of Maunakea has always had within the indigenous Hawaiian community.  We are most fortunate to have the opportunity to conduct observations from this mountain. 
\end{acknowledgements}

\end{document}